\documentclass[aps,pre, twocolumn,showpacs,epsfig,amsmath,amssymb,eqsecnum]{revtex4}
\usepackage{bbm}
\usepackage{mathrsfs}
\usepackage{amsfonts}
\usepackage{color}
\usepackage[dvips]{graphicx}% Include figure files
\usepackage{bbm}
\usepackage{amsfonts}
\usepackage{color}
\newcommand{\xv}{{\vec x}}

\newcommand{\Tr}{{\rm Tr}}

\newcommand{\llm}{{\boldsymbol{\lambda}}}

\newcommand{\be}{\begin{equation}}
\newcommand{\ee}{\end{equation}}
\newcommand{\bea}{\begin{eqnarray}}
\newcommand{\eea}{\end{eqnarray}}
\begin{document}
\title{Biaxial Deformations of Rubber: Entanglements or Elastic Fluctuations?}
%\preprint{APS/123-QED}
\author{Xiangjun Xing}
\address{Institute of Natural Science and Department of Physics, 
Shanghai Jiao Tong University, Shanghai, 200240, China}
%Syracuse University, Syracuse, New York~13244}
\email{xxing@sjtu.edu.cn}
%\author{Kenji Urayama}
%\affiliation{Department of Materials Chemistry, Kyoto University, Kyoto 615-8510, Japan}
% \date{\today} %freeze this upon submission
% but any date may be explicitly specifie

\begin{abstract} 
The classical theory of rubber elasticity fails in the regime of large deformation.  The underlying physical mechanism has been under debate for long time.  In this work, we test the recently proposed mechanism of thermal elastic fluctuations  by Xing, Goldbart and Radzihovsky~\cite{XGR-rubber} against the biaxial stress-strain data of three distinct polymer networks with very different network structures, synthesized by Urayama~\cite{Urayama-biaxial-2}  and Kawabata~\cite{Kawabata:1981ao} respectively.  We find that both the two parameters version and the one-parameter version of the XGR theory provide satisfactory description of the elasticity in whole deformation range.  For comparison, we also fit the same sets of data using the slip-link model by Edwards and Vilgis with four parameters.  The fitting qualities of two theories are found to be comparable.  
%Based on this result, we argue that after the effects of elastic fluctuations taken into account, there is no mechanical evidence supporting entanglement theories.   

%xing.1972@gmail.com

%A recent theory of rubber elasticity by Xing, Goldbart and Radzihovsky~\cite{XGR-rubber} attributes the failure of the classical rubber elasticity theory to the entropy of long wavelength elastic fluctuations, as supposed to the entanglement effects.   In this work, we show that this theory successfully explains the general biaxial stress-strain data of two polymer networks with very different network structures, synthesized by Urayama~\cite{Urayama-biaxial-2}  and Kawabata~\cite{Kawabata:1981ao} respectively.  We argue that after the effects of elastic fluctuations taken into account, there is no mechanical evidence supporting entanglement theories.   

%This provides further support for a 
\end{abstract}

% [CHECK: need to add PACS numbers]

% PACS, the 
%61.41.+e 61.43.-j
%Physics and Astronomy
\maketitle

 \section{Introduction}
The classical theory of rubber elasticity~\cite{EL:Treloar}  properly captures the essence of entropic elasticity of polymer networks.  As it is well known, however, it fails to predict accurately the stress-strain relation at large deformation regime.  Understanding of the mechanism of this failure and attempting to improve it has remained a central problem in field of polymer science in the last half century. 

The classical theory is based on two basic assumptions: 1) The probability distribution of the end-to-end displacement of polymer chains are Gaussian; 2) the total entropy of the network is the sum of entropy of each chain.  Regarding the assumption 1), while it is well known that polymer chains are Gaussian, i.e. ideal, in a polymer melt, wether this remains true after the system being crosslinked is not so clear.  It is however the fundamental assumption of almost all entanglement theories that this assumption is incorrect due to topological constraints imposed by entanglements.  Exact treatment of entanglement effects is strictly speaking a complicated topological problem, and does not seem possible at this stage.  Hence the possibility that entanglements do not invalidate the classical theory in the limit of long chain can not be excluded.  Nevertheless, different researches have made distinct assumptions about the effects of entanglements in rubber elasticity, and have developed different versions of entanglement theory.  These include the Slip-Link Model by Edwards and Vilgis \cite{Edwards-tube-1986}, the diffused-constraint model  by Kloczkowski, Mark, and Erman \cite{Kloczkowski:1995lr}, the localization model by Gaylord and Douglas \cite{Gaylord-Douglas-I,Gaylord-Douglas-II}, as well as the slip-tube model by Rubinstein and Panyukov \cite{Rubinstein-tube}.   
%:
%As These models 
%:

The assumption 2) seems much less challenged, but can be easily seen to be invalid.   It is obvious that the entropy of network junctions is nonzero and is not included in the classical theory.  James and Guth~\cite{james-guth} has shown that, for phantom networks, this part of entropy is independent of strain deformation, and therefore has no effect on the elasticity.   Historically, this result may has been seen as an evidence supporting entanglement theories.  
%
%Note, however, Kloczkowski, Mark, and Erman have considered junction fluctuations in the presence of entanglement in their model.   \xing{More discussion is needed here. }
%have the important consequence of diverting the attention to the challenge of the assumption 1).   
%\xing{Need more discussion on other variants of entanglement theories. Check with kenji's papers.  Especially his most recent paper on this topic!} However, some theories have considered the role of junction fluctuations approximately.  ...
%
Real rubbery materials are, however, nearly incompressible, with Poisson ratio $1/2$.  Junction fluctuations are essentially different in phantom networks and in incompressible networks.  This difference was made clear in a recent work~\cite{XGR-rubber} by Xing, Goldbart, and Radzihovsky (hereafter referred as XGR).  At long length scale, these fluctuations can be simply modeled as thermal elastic fluctuations, i.e. phonon fluctuations, of an isotropic solid.  In incompressible solids, longitudinal phonons are infinitely massive and therefore do not renormalize the elasticity.  It turns out that the fluctuations of transverse phonon modes depend sensitively on the macroscopic strain deformation, and are quantitatively comparable to the effect of single chain entropy addressed by the classical theory.  
%:
%The calculated stress-strain curved of uniaxial deformation, with these fluctuations included at lowest order, provide a good fit to the experimental data for uniaxial deformations.  
%:
This strong interplay between long wave length thermal fluctuations and elastic properties is not peculiar to rubber at all.  In fact it has been one of the main theme of modern condensed matter physics, and has been analyzed in many different systems, such as polymers, membranes, and liquid crystals.   

Both entanglement theories and XGR theory fit well to uniaxial stress-strain data.  To distinguish these theories, therefore, it is important to test them aganist data of more general biaxial deformations.  \footnote{Experimental studies of chain conformation and junction fluctuations at single-molecular level may also provide useful information, but we do not know any relevant systematic experimental data at hand. }  Recently Urayama {\it et. al.}~\cite{Urayama-biaxial-2} tested five different versions of entanglement theories using biaxial stress-strain data of a well-characterized end-linked networks.  The slip-link model by Edwards and Vilgis, hereafter referred as EV, was found to provide the best fit.  In a more recent work\cite{PhysRevE.77.011802}, Hansen, Skov, and Hassager tested XGR theory and a version of slip-link model against the unaxial stress-strain data by Mark {\it et al}, and concluded that XGR and slip-link model fit the uniaxial data with comparable quality.  In this work, we use two sets of biaxial data by Urayama, as well as another set of earlier data by Kawabata {\it et. al.} \cite{Kawabata:1981ao}  to test the XGR's elastic fluctuation theory, and compare it with EV's slip-link theory.

\section{Theories}
%{\bf Theories} \quad 
Let an incompressible isotropic solid be biaxial deformed.  A mass point $\xv$ is deformed to $\xv' = \llm \cdot \xv$, where the deformation gradient matrix $\llm$ has three eigenvalues $\lambda_1,\lambda_2,\lambda_3$ in three orthogonal directions.  The incompressibility constraint imposes the relation: $\lambda_3 = 1/(\lambda_1 \lambda_2)$; therefore only two of these three variable can be independently controlled. 
%and the associated metric tensor ${\mathbf g}$ is given by
%\be \llm =  \begin{pmatrix}
%\lambda_1 & 0& 0\\
%0 & \lambda_2 & 0\\
%0& 0& \left( \lambda_1 \lambda_2 \right)^{-1}
%\end{pmatrix}, \quad 
%{\mathbf g} =  \llm^{\rm T} \llm = 
%\begin{pmatrix}
%\lambda_1^2 & 0& 0\\
%0 & \lambda_2^2 & 0\\
%0& 0& \left( \lambda_1 \lambda_2 \right)^{-2}
%\end{pmatrix}. \ee
%:
%Note that we have taken into account the incompressibility by construction of $\llm$ and ${\mathbf g}$.   
%:
If we impose deformations $\lambda_1,\lambda_2$ along two axes and let the third axes relax freely, the two components of nominal (engineer) stress (measured relative to the undeformed area) are given by \cite{EL:Treloar}
\be
\sigma_1 = \frac{\partial f}{\partial \lambda_1},
\quad 
\sigma_2 = \frac{\partial f}{\partial \lambda_2},
\label{stress-biaxial-def}
\ee
with $\lambda_3$ treated as a function of $\lambda_1,\lambda_2$.  The nominal stress $\sigma_3$ along the third axis is zero, since the third dimension can relax freely.

%\subsection{XGR Theory}

According to the XGR theory, the total elastic free energy density is characterized by two tunable parameters: 
\be
f (\llm) = \frac{\mu_0}{2}\, \Tr \, {\mathbf g} 
+ \mu_1 \, \langle \log \hat{q} \cdot {\mathbf g}^{-1} \cdot  \hat{q} \rangle_{\hat{q}} .  
\label{XGR-free_energy}
\ee 
Here the average $\langle \,\, \cdot  \,\, \rangle_{\hat{q}}$ is over the orientation of the 3d unit vector $\hat{q}$, and ${\mathbf g} = \llm^{\rm T} \llm $ is called the {\em metric tensor}.  The first term is just the classical rubber elasticity theory and is due to the entropy of polymer chains.  The parameter $\mu_0$ has the physical interpretation of the effective density of polymer chains that contributing to the network elasticity.  This value however may receive substantial contribution from entanglement effects, and therefore may be significantly different from the real chain density.   \footnote{ The idea that entanglements act effectively like crosslinking and enhances the shear modulus is well accepted and is supported by experiments and simulations.  What is questionable is whether entanglements are responsible for the failure of the classical theory in the large deformation regime. }  The second term is the lowest order correction due to elastic fluctuations; $\mu_1$ is half the total number of phonon modes in unit volume.  Two parameters $\mu_0,\mu_1$ are therefore expected to be of the same order of magnitude.  In reality, however, their best-fitting values are always found to be very close.  There maybe relevant physics underlying this intriguing coincidence.

%For uniaxial deformation $\lambda_1 = \lambda, \lambda_2 = 1/\sqrt{\lambda}$, the angular average in the second term can be analytically calculated.  The total elastic free energy is then given by \cite{XGR-rubber}: 
%\begin{eqnarray} \hspace{-0.3cm}
%f =  \frac{a}{2} \, (\lambda^2 + \frac{2}{\lambda})
%+2\, b \left[  \frac{\tanh^{-1}\sqrt{1-\lambda^{-3}}}{\sqrt{1-\lambda^{-3}}}
 %- \ln\lambda \right].  \label{elastic-f-2} \end{eqnarray}
%The nominal stress components can also be calculated accordingly (assuming that the system is stretched along the direction of $\hat{e}_1$):
%\be  \sigma_1 = \frac{\partial f}{\partial \lambda_1}, \quad
%\sigma_2 = \sigma_3 = 0. \ee
%For uniaxial compression, $\lambda < 1$, and the above expression needs to be analytically continued appropriately.  For biaxial deformation, however, the angular average has to be calculated numerically.  

%\subsection{Slip-link Theory (EV)}

The slip-link model by Edwards and Vilgis \cite{Edwards-tube-1986} models the effect of entanglement between two chains as a slip-link.  The free energy density \footnote{We use the result presented in the original paper by Edwards and Vilgis \cite{Edwards-tube-1986}. } is characterized by four parameters:
\begin{subequations}
\bea
f &=& \frac{1}{2} C_1 \left[ \frac{
   \left(1-\alpha ^2\right) S}{1- \alpha ^2S } 
   - \log (1-\alpha^2 S )\right] 
     \nonumber\\
 &+&
 \frac{1}{2} C_2  \left[ \sum_{i = 1}^3
 \frac{(1-\alpha^2 ) (1+ \eta ) \lambda_i^2}
   {\left(1- \alpha^2 S \right) \left(1+ \eta 
   \lambda_i^2 \right)}
  \right.
  \nonumber\\
 &- &   \left.   \log (1- \alpha^2 S )
 +  \log \prod_{i = 1}^3 \left(1+ 
 \eta \lambda_i^2 \right)
     \right],   
\eea
     \label{EV-free_energy}
\end{subequations}
where $S = \lambda_1^2 + \lambda_2^2 + \lambda_3^2$, $\eta$ is the slippage parameter, being zero if the entanglement behaves completely as a cross-link; the parameter $\alpha$ characterizes the finite extensibility of polymer chains.   $C_1$ is the polymer chain density, while $C_2$ is the slip-link density.   This theory reduces to the classical theory when both $\eta,\alpha$ vanish.  

\section{Experimental Systems}
The two systems synthesized by Urayama {\it et. al.} \cite{Urayama-biaxial-1} are end-linked poly(dimethylsiloxane) (PDMS) networks using tetrafunctional crosslinker tetrakisdimethylsiloxysilane. The number- and weight-average molecular weights (Mn and Mw, respectively) of the precursor PDMS were evaluated to be 46 600 and 89 500, respectively, by gel permeation chromatography (GPC).  In the first system, no diluent is added in the whole process; the resulting polymer network shall be referred as $100\%$ sample.  In the second system, $30\%$ of diluent is added in the polymerization process and is kept afterwards.  The resulting system shall be referred as the $70\%$ sample.  All experiments were done at the temperature of $40^oC$.  The physical chain densities were estimated to be $5.88 mol/m^3$ for the $100\%$ sample and $3.22 mol/m^3$ for the $70\%$ sample.  \cite{Urayama-biaxial-4}  The system prepared by Kawabata is a vulcanized isoprene rubber (IR) with polydipersed distribution of chain length.  No structure data is known from the preparation method.  

%Our fitting strategy is as follows: we first identify all the data corresponding to uniaxial deformations.  We use the XGR theory to fit these data, and extract two parameters which completely fix the theory.  We then use the theory to calculate stress-strain relation of biaxial deformations and compare them with the biaxial data.   A successful prediction of stress-strain curves for biaxial deformations shall constitute a strong justification of the XGR theory.  

\section{Data Fitting}

%{\bf Data fitting} \,\, 
%%:
Fitting of stress-strain data using XGR theory is simple.   From Eq.~(\ref{XGR-free_energy}) the stress components is linear in $\mu_0,\mu_1$.  The variance is therefore a positive-definite quadratic function of these two parameters:
\be
\sum_{i} \big [ 
\mu_0 \, \sigma_0(i) + \mu_1 \, \sigma_1(i) - \sigma_{\rm exp}(i)
\big ]^2,
\ee
where the sum runs over all data points.  Minimization of this variance is a trivial matter,  since it is guaranteed that there is only one minimum.  

By contrast, fitting of stress-strain data using EV theory is more complicated.  The slip-link theory has four parameters, $C_1,C_2, \eta,\alpha$, see 
\ref{EV-free_energy}.  Among these, $\eta$ and $\alpha$ enter the theory in an essentially nonlinear way.   We found that the variance as a function of four parameters $C_1,C_2,\eta,\alpha$ may exhibit multiple local minima that are approximately degenerate.  In this case, the parameters will not be able to be determined using stress-strain data alone.  For Urayama's systems, however, the parameter $C_1$ in Eq.~(\ref{EV-free_energy}) can be determined \cite{Urayama-biaxial-2,Urayama-biaxial-4} separately using structure information.    

\subsection{Urayama's $100\%$ Sample}
\begin{figure}
\begin{center}
\includegraphics[height=4cm]{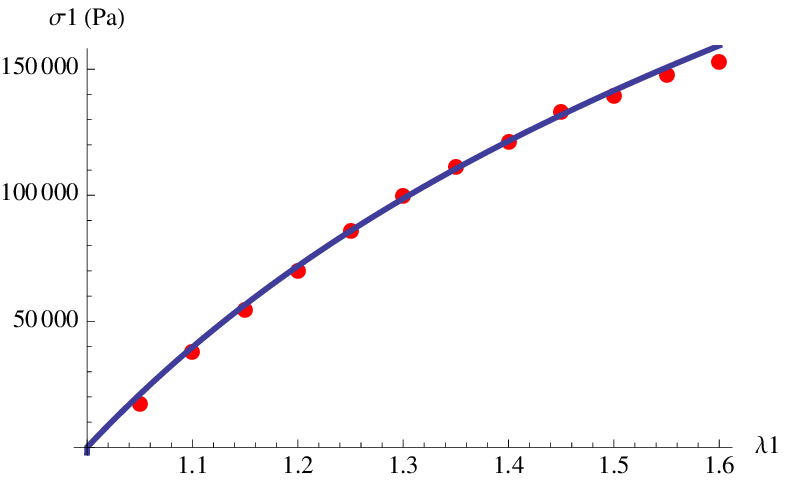}
\caption{XGR fitting of the uniaxial stress-strain data of  Urayama's $100\%$ sample \cite{Urayama-biaxial-2}.  }
\label{uniaxial-fitt-Urayama}
\end{center}
%\vspace{-5mm}
\end{figure}

\begin{figure}[htb]
\begin{center}
\includegraphics[height=5cm]{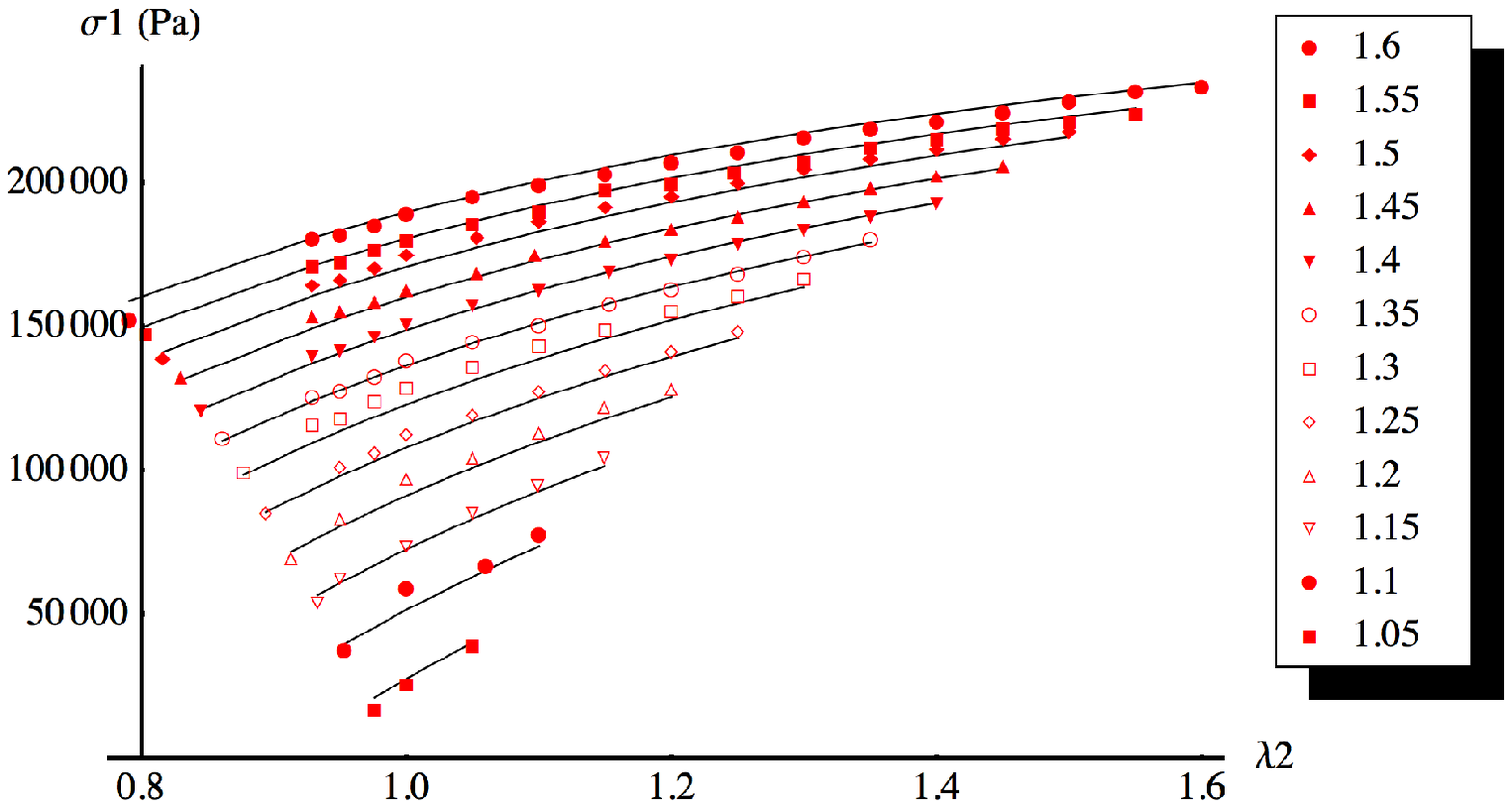}
\caption{XGR fitting of stress component $\sigma_1 = \partial f/\partial \lambda_1$ for biaxial deformation of Urayama's $100\%$ sample.  Each symbol corresponds to a distinct value of $\lambda_1$.  }
\label{biaxial-fitt-Urayama-1}
\end{center}
%\vspace{-5mm}
\end{figure}

\begin{figure}
\begin{center}
\includegraphics[height=5cm]{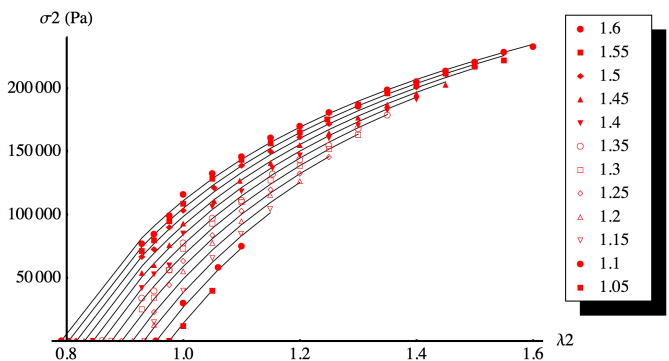}
\caption{XGR fitting of stress component $\sigma_2 = \partial f/\partial \lambda_2$ for biaxial deformation of Urayama's $100\%$ sample.  Each symbol corresponds to a distinct value of $\lambda_1$.   }
\label{biaxial-fitt-Urayama-2}
\end{center}
%\vspace{-5mm}
\end{figure}

We first fit Urayama's data for the $100\%$ sample \cite{Urayama-biaxial-2}  using XGR theory.  The best-fitting parameters for $\mu_0,\mu_1$ are very close to each other, as shown in the first row of Table \ref{table-Urayama}. The fitting of uniaxial stress-strain data is shown in Fig.~\ref{uniaxial-fitt-Urayama}.  The fitting of two stress components $\sigma_1$ and $\sigma_2$ for general biaxial data are shown in Fig.~\ref{biaxial-fitt-Urayama-1} and Fig.~\ref{biaxial-fitt-Urayama-2} respectively.     We also perform the data-fitting with the constraint $\mu_0 = \mu_1$.   As shown in the second row of Table \ref{table-Urayama}, the resulting variance is almost the same as the unconstrained two parameter fitting. Furthermore, the stress strain curves of two cases show no noticeable difference.   This points to the interesting possibility that their theoretical values are indeed the same, for some reason to be understood.   

Let us define an effective chain density $n^{eff}_c$ via 
\be
\mu_0 = n^{eff}_c k_B T, 
\ee
where the temperature $T = 40^oC = 313K $, see  \cite{Urayama-biaxial-2,Urayama-biaxial-4}.  The best fitting value for $\mu_0$ shown in Table \ref{table-Urayama} then translates into an effective chain density $n^{eff}_c = 40.0 mol/m^3$, which is much larger than the physical chain density, estimated to be $n_c = 5.88 mol/m^3$\cite{Urayama-biaxial-2,Urayama-biaxial-4}.  The parameter $\mu_0$ in XGR theory receives most contributions from entanglements, rather than from chemical cross-links.    

\begin{widetext}
\begin{center}
\begin{table}[!hbt]
%%\vspace{5mm}
\begin{tabular}{|c|c|c|}
\hline\hline
Theory & Parameters ($MPa$) & Variance ($10^9 Pa^2$)
\\ \hline
XGR ($\mu_0 \neq \mu_1$) & $\mu_0 = 0.104, \mu_1=0.112$ & $2.745 $ 
\\\hline
XGR ($\mu_0 = \mu_1$) & $\mu_0 = \mu_1=0.106$ & $2.805 $ 
\\\hline
EV (unrestricted) & $(-0.0019, 0.165, 0.120,  0.155)$
& $2.481$
\\\hline
EV (restricted)\footnote{The parameter $C_1$ is determined independently.  The other three parameters are determined by data fitting.} & $(0.0152, 0.150, 0.143, 0.152)$
& $2.483$
\\\hline\end{tabular}
\caption{Best-fitting parameters and variances of Urayama's $100\%$ sample \cite{Urayama-biaxial-2}. The four parameters in EV theory are $(C_1,C_2,\eta,\alpha)$ respectively. }
\label{table-Urayama}
\end{table}
\end{center}
\end{widetext}

\begin{figure}
\begin{center}
\includegraphics[height=5cm]{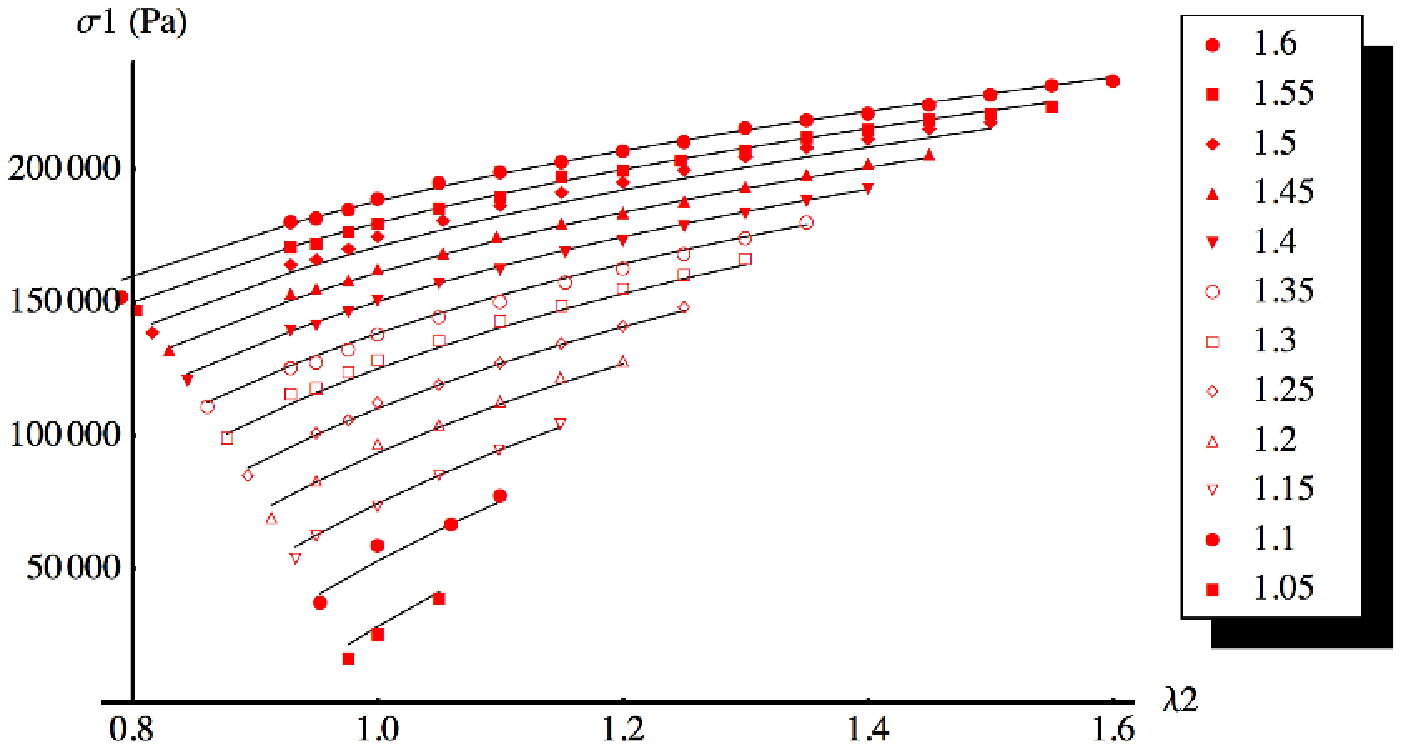}
\caption{EV fitting of stress component $\sigma_1 = \partial f/\partial \lambda_1$ for biaxial deformation of Urayama's $100\%$ sample.  Each symbol corresponds to a distinct value of $\lambda_1$.  }
\label{biaxial-fitt-Urayama-1-EV}
\end{center}
%\vspace{-5mm}
\end{figure}

\begin{figure}
\begin{center}
\includegraphics[height=5cm]{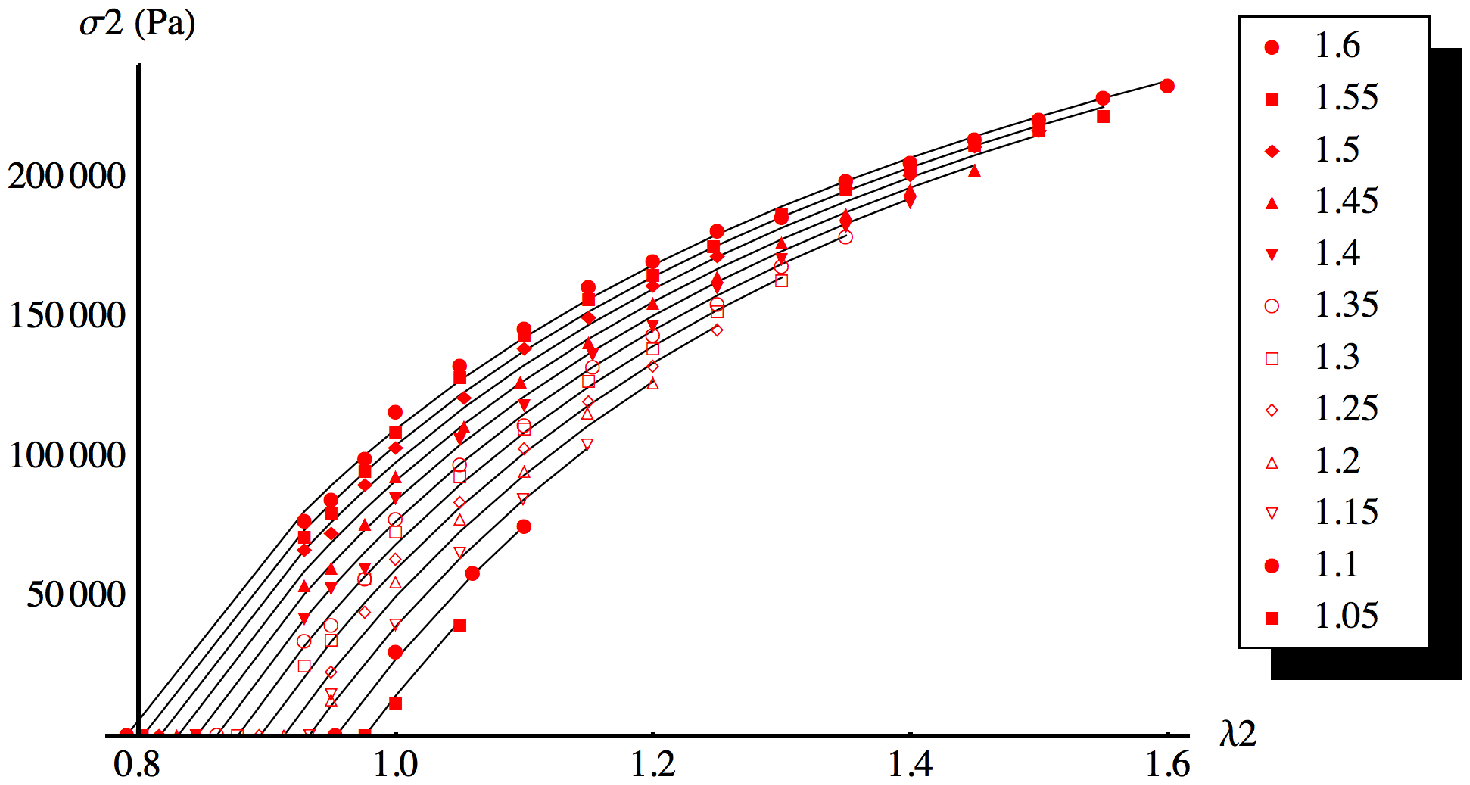}
\caption{EV fitting of stress component $\sigma_2 = \partial f/\partial \lambda_2$ for biaxial deformation of Urayama's $100\%$ sample.  Each symbol corresponds to a distinct value of $\lambda_1$.   }
\label{biaxial-fitt-Urayama-2-EV}
\end{center}
%\vspace{-5mm}
\end{figure}

We also fit the same set of data using EV theory.   As shown inEq.~(\ref{table-Urayama}), an unrestricted fitting with all four parameters free to change yields unphysical results: the best-fitting parameter $\mu_0$ is negative.   
%:
%%We note in reference \cite{Urayama-biaxial-2}, the EV theory was expanded into series of $\alpha$ before being used for fitting.  It was found there that the best fitting parameters are physical.   (\xing{Kenji needs to verify this.})  
%
Urayama \cite{Urayama-biaxial-4} independently determined the value of $C_1$ to be $ 0.0152 MPa$ using the knowledge of physical chain density.  We have also followed the same procedure.   The fitting quality thus obtained is almost identical to the unrestricted data fitting.   The restricted fitting results of biaxial stress-strain data using EV theory are shown in Eq.~(\ref{biaxial-fitt-Urayama-1-EV}) and Eq.~(\ref{biaxial-fitt-Urayama-2-EV}) respectively.   We have checked the individual stress-strain data curves, and find no noticeable difference between the restricted fitting and the unrestricted fitting.  

% This insensitivity of fitting quality to the change of $C_1$ shows that, at least for Urayama's $100\%$ sample, it is not possible to determine all the parameters in EV theory using stress-strain data only.  This feature is rather easy to understand.  In the EV theoryEq.~(\ref{EV-free_energy}, the second term is due to slip links.  If the slippage parameter $\eta$ vanishes, a slip link becomes identical to a cross-link, and two terms inEq.~(\ref{EV-free_energy} becomes identical to each other.  It becomes therefore impossible to determine two parameters separately by elastic experiments.   In reality, the best fitting value for $\eta$ turns out to be small, around $0.12 \sim 0.14$, seeEq.~(\ref{table-Urayama}.  Hence the contributions of chemical crosslinks is very similar to the those of entanglement.  Change of the parameter $C_1$ (crosslink density) can be compensated by change of $C_2$ (slip link density) to very high precision.  
%This explains the insensitivity of the fitting quality to the change of $\mu_0$.   Technically, however this implies that we are not able to determine the ``physical values'' of all parameters from stress-strain data up to reasonable precision:  The slip-link theory becomes pathological in the regime of small slippage $\eta$.  

\subsection{Urayama's $70\%$ sample}
\begin{figure}
\begin{center}
\includegraphics[height=4cm]{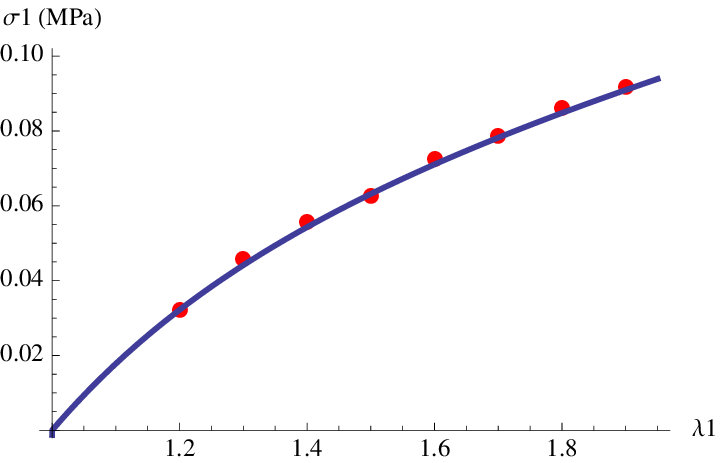}
\caption{XGR fitting of the uniaxial stress-strain data of  Urayama's $70\%$ sample \cite{Urayama-biaxial-2}.  }
\label{uniaxial-fitt-Urayama-70}
\end{center}
%\vspace{-5mm}
\end{figure}

\begin{figure}
\begin{center}
\includegraphics[height=5cm]{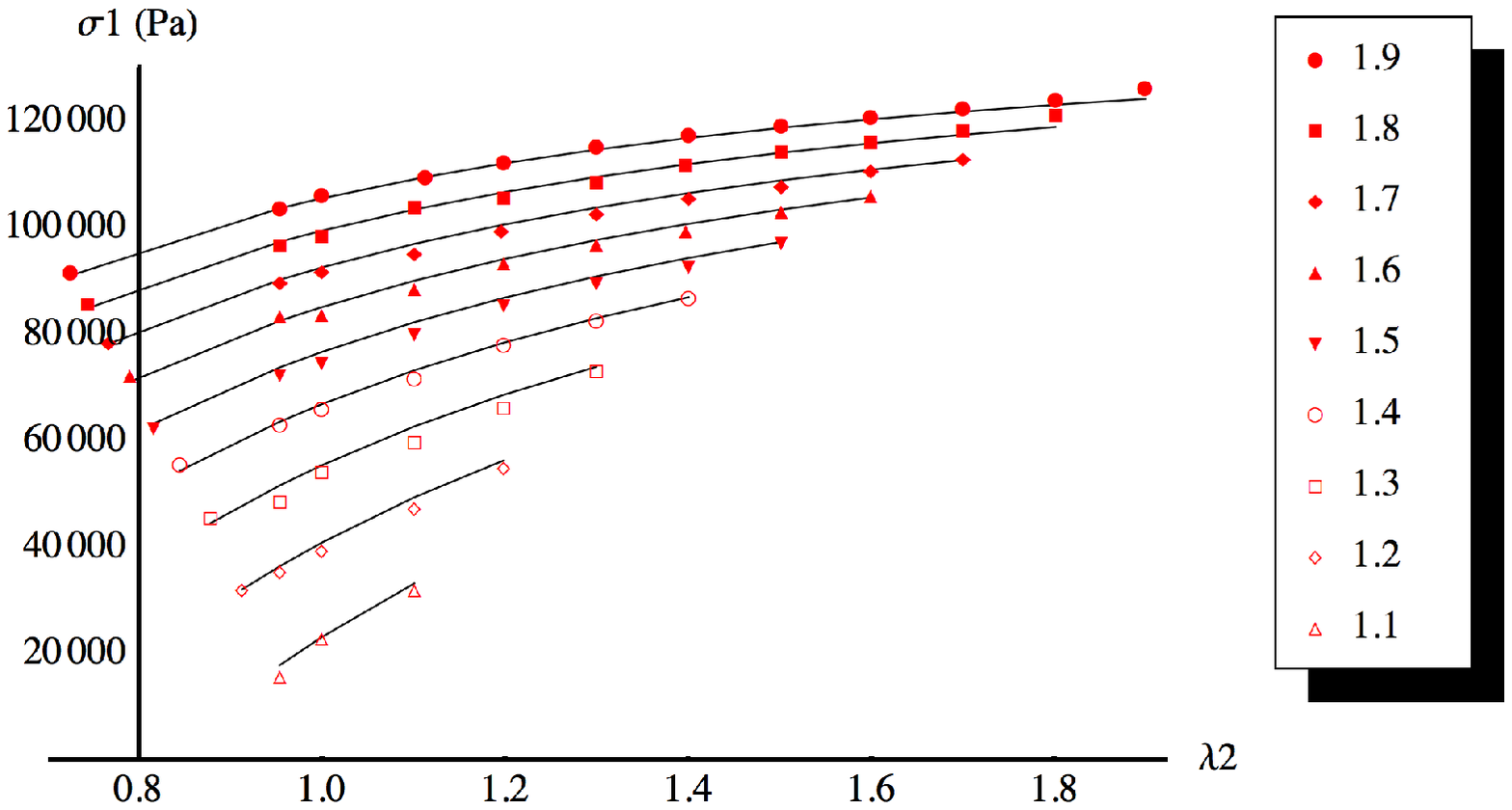}
\caption{ XGR fitting of stress component $\sigma_1 = \partial f/\partial \lambda_1$ for biaxial deformation of Urayama's $70\%$ sample.  Each symbol corresponds to a distinct value of $\lambda_1$.}
%Experimental data and theoretical prediction for $\sigma_1$ as function of $\lambda_2$ for given $\lambda_1$.   The left most points of all branches correspond to uniaxial deformation and were used to extract two parameters $a,b$. }
\label{biaxial-fitt-Urayama-1-70}
\end{center}
%\vspace{-5mm}
\end{figure}

\begin{figure}
\begin{center}
\includegraphics[height=5cm]{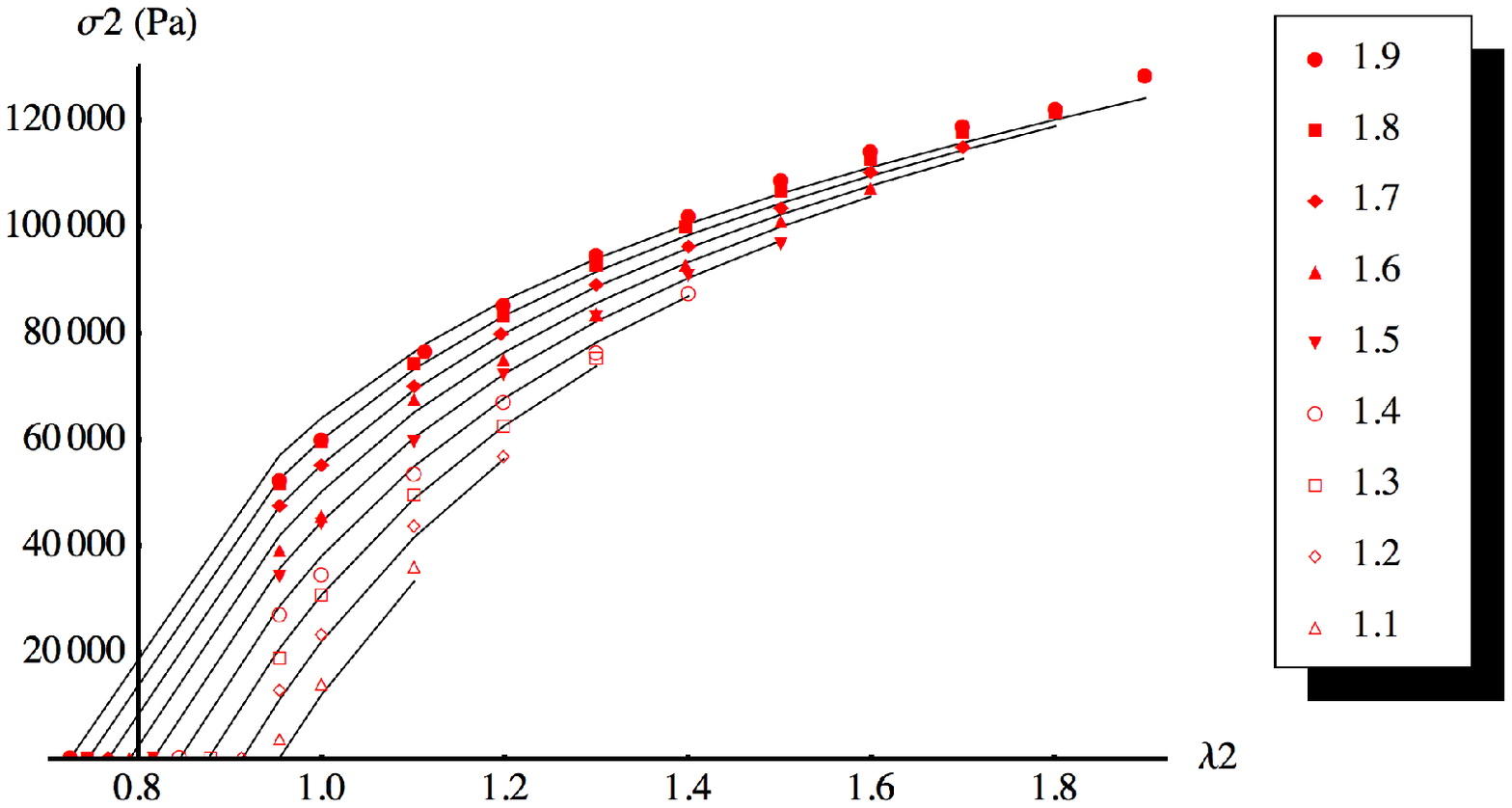}
\caption{
XGR fitting of stress component $\sigma_2 = \partial f/\partial \lambda_2$ for biaxial deformation of Urayama's $100\%$ sample.  Each symbol corresponds to a distinct value of $\lambda_1$. }
%Experimental data and theoretical prediction for $\sigma_2$ as function of $\lambda_2$ for given $\lambda_1$. }
\label{biaxial-fitt-Urayama-2-70}
\end{center}
%\vspace{-5mm}
\end{figure}

We have also fitted Urayama's $70\%$ sample using the same procedure.  This system is prepared in $30\%$ (weight) concentration of good solvent.   The solvent is kept in the measurement stage.   As in the $100\%$ sample, the best fitting parameters for $\mu_0,\mu_1$ in the XGR theory are again very close to each other. 
The value of $\mu_0$ translates into an effective chain density $n^{eff}_c \approx 17.5 mol/m^3$, which is much larger than the upper bound of real chain density $n_c = 3.22 mol/m^3$, determined independently by Urayama \cite{Urayama-biaxial-2,Urayama-biaxial-4}.  Hence $\mu_0$ is again dominated by entanglement effects.  The fitting results of uniaxial and biaxial stress-strain data are shown in Eq.~(\ref{uniaxial-fitt-Urayama-70}), Eq.~(\ref{biaxial-fitt-Urayama-1-70})  and Eq.~(\ref{biaxial-fitt-Urayama-2-70}) respectively.
 
The unrestricted fitting using EV slip-link theory yields parameter values 
that are all physical.  The restricted fitting, with the parameter $C_1$ fixed independently using structure data, yields substantially different result for the slippage parameter $\eta$, and a total variance that is very close to that of XGR fitting.  This is probably a generic feature of nonlinear data fitting.  
%This substantial difference between the restricted and the unrestricted fittings is a discomforting feature of the EV theory.  

Caution should be taken in the fitting of either theory to stress-strain data of swollen network.   Both EV theory and XGR theory are improvements over the classical theory of rubber elasticity, which corresponds to the first term of 
\ref{XGR-free_energy} and Eq.~(\ref{EV-free_energy}).  The classical theory, however, constitutes a zero-th order approximation only for {\em un-swollen} polymer networks, but not for swollen networks.  The first basic assumption, that polymer chains obey Gaussian statistics, breaks down when chains are swollen by good solvent.  It is well known that polymers in good solvent obey very different, nonGaussian statistics \cite{polymer:deGennes}.   Indeed the fitting quality of both theories for the $70\%$ sample becomes noticeably worse than that of 
the $100\%$ sample in the large deformation regime.   We have tested that the fitting quality becomes even worse for samples with higher degree of swelling.

For comparison, we also show the restricted fitting results of biaxial stress-strain data using EV theory in Eq.~(\ref{biaxial-fitt-Urayama-1-EV}) and Eq.~(\ref{biaxial-fitt-Urayama-2-EV}) respectively.  Again no substantial difference in fitting quality can be detected between fittings using two theories.  

\begin{figure}
\begin{center}
\includegraphics[height=5cm]{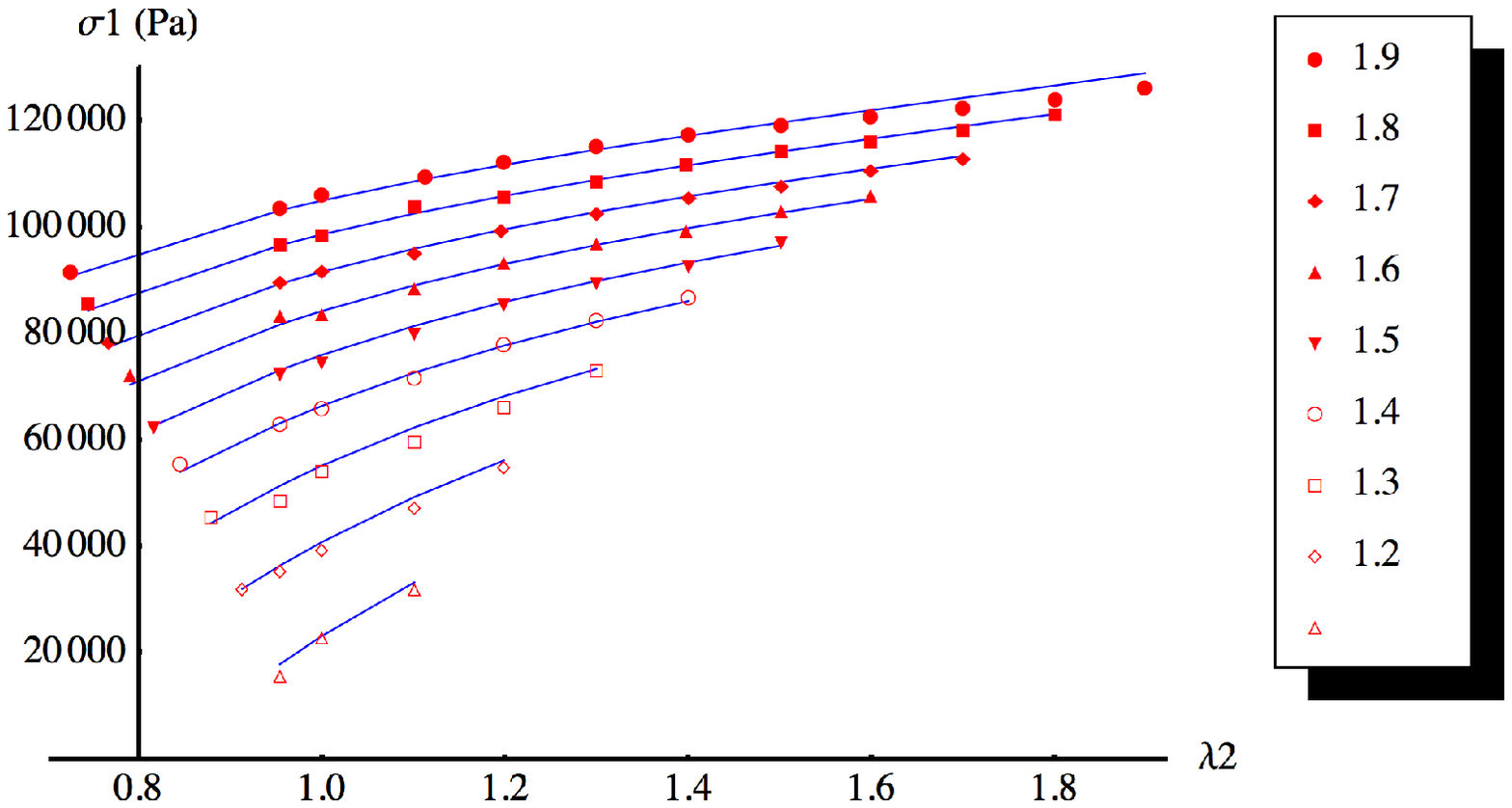}
\caption{Restricted EV fitting of stress component $\sigma_1 = \partial f/\partial \lambda_1$ for biaxial deformation of Urayama's $70\%$ sample.  Each symbol corresponds to a distinct value of $\lambda_1$. }
% Restricted fitting of stress component $\sigma_1 = \partial f/\partial \lambda_1$ for biaxial deformation of Urayama's $70\%$ sample using EV's slip-link theory.  Each symbol corresponds to a distinct value of $\lambda_1$.  }
\label{biaxial-fitt-Urayama-1-70-EV}
\end{center}
%\vspace{-5mm}
\end{figure}

\begin{figure}
\begin{center}
\includegraphics[height=5cm]{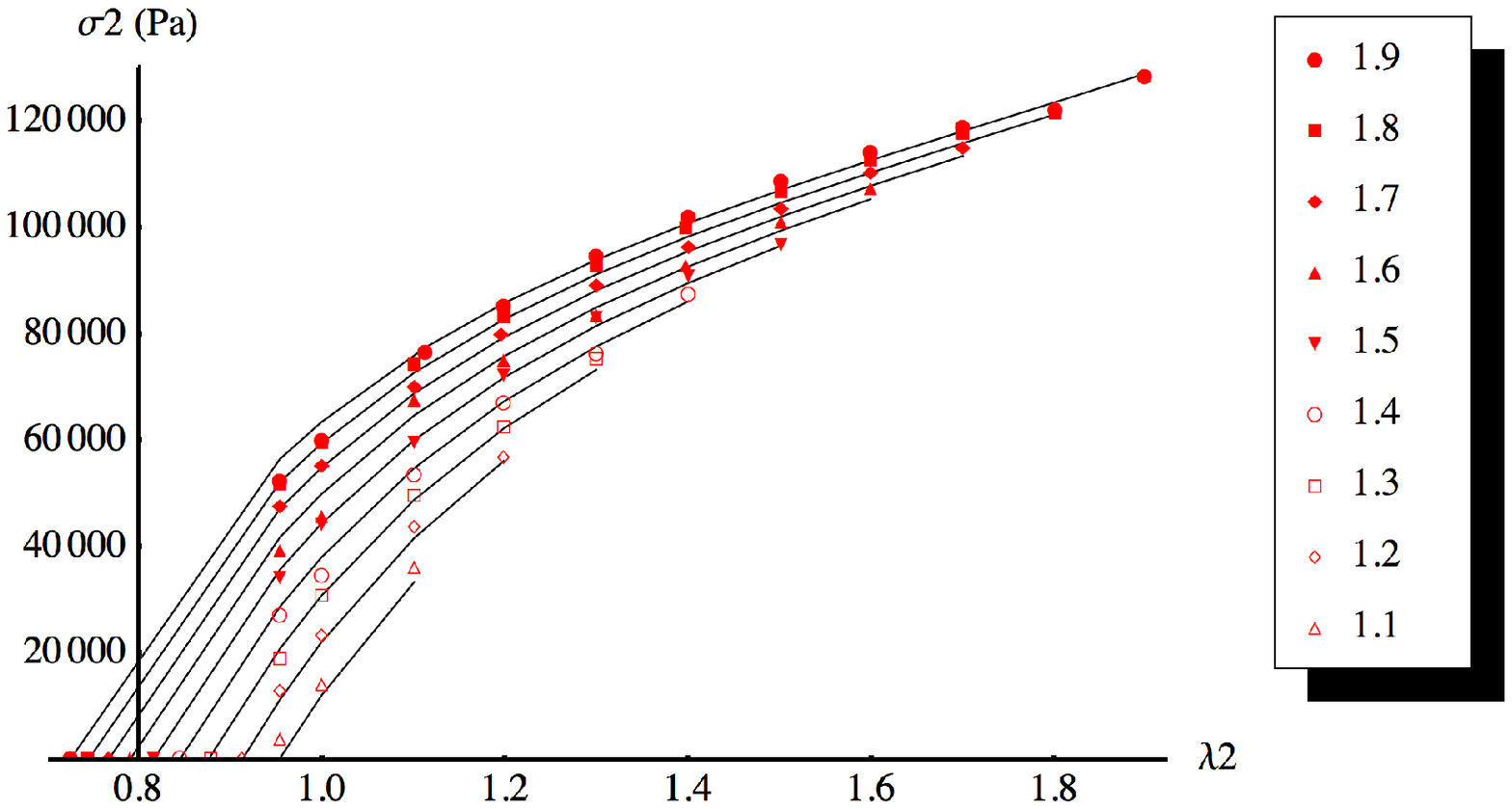}
\caption{
Restricted EV fitting of stress component $\sigma_2 = \partial f/\partial \lambda_2$ for biaxial deformation of Urayama's $70\%$ sample.  Each symbol corresponds to a distinct value of $\lambda_1$. }
%Restricted fitting of stress component $\sigma_2 = \partial f/\partial \lambda_2$ for biaxial deformation of Urayama's $70\%$ sample using EV's slip-link theory.  Each symbol corresponds to a distinct value of $\lambda_1$.   }
\label{biaxial-fitt-Urayama-2-70-EV}
\end{center}
%\vspace{-5mm}
\end{figure}

\begin{widetext}
\begin{center}
\begin{table}[!hbt]
%\vspace{5mm}
\begin{tabular}{|c|c|c|}
\hline\hline
Theory & Parameters ($MPa$) & Variance ($10^9 Pa^2$)
\\ \hline
XGR ($\mu_0 \neq \mu_1$) & $\mu_0 = 0.045, \mu_1=0.053$ & $0.369 $ 
\\\hline
XGR ($\mu_0 = \mu_1$) & $\mu_0 = \mu_1=0.048$ & $0.442 $ 
\\\hline
EV (unrestricted) 
& $(0.032, 0.052, 0.419, 0.134)$
& $0.282$
\\\hline
EV (restricted)\footnote{The parameter $C_1$ is determined independently.  The other three parameters are determined by data fitting.} 
& $(0.0083, 0.064, 0.138,  0.155)$
& $0.354$
\\\hline\end{tabular}
\caption{Best-fitting parameters and variances of Urayama's $70\%$ sample \cite{Urayama-biaxial-2}. The four parameters in EV theory are $(C_1,C_2,\eta,\alpha)$ respectively.
The fitting result using EV theory is slightly different from that by Urayama \cite{Urayama-biaxial-2}.  }
\label{table-Urayama-70}
\end{table}
\end{center}
\end{widetext}

\subsection{Kawabata's Uncharacterized Sample}
We also use XGR and EV to fit Kawabata's data \cite{Kawabata:1981ao}, measured using network with much greater randomness.  The results for best-fitting parameters are shown in Eq.~(\ref{table-Kawabata}).  Again for the two parameter version of XGR theory, $\mu_0,\mu_1$ are very close to each other.   The one parameter version of XGR (with $\mu_0=\mu_1$)
leads to $30\%$ more variance, but the stress-strain curves hardly show any visual difference.  The best fitting parameters for EV theory are completely physical.  Unlike Uarayama's sample, there is no available structure information on Kawabata's polymer network; the parameter $C_1$ therefore can not be determined independently.   An unrestricted fitting using EV theory yields slightly better fitting quality, with $25\%$ less variance then XGR theory.  The difference is however not substantial, given the fact that EV theory has two more parameters than XGR.  The fitting result of Kawabata's data using EV theory is rather robust against variations of initial conditions.  This is likely related to the fact that the best-fitting value for the slippage parameter $\eta \sim 0.31662$ is much larger that of Urayama's samples.   

\begin{figure}
\begin{center}
\includegraphics[height=5cm]{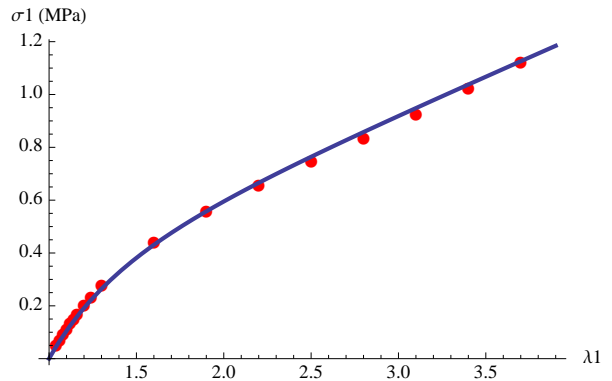}
\caption{XGR fitting of the uniaxial stress-strain data of  Kawabata's sample \cite{Kawabata:1981ao}. }
\label{uniaxial-fitt-Kawabata}
\end{center}
%\vspace{-5mm}
\end{figure}

\begin{figure}
\begin{center}
\includegraphics[height=5.5cm]{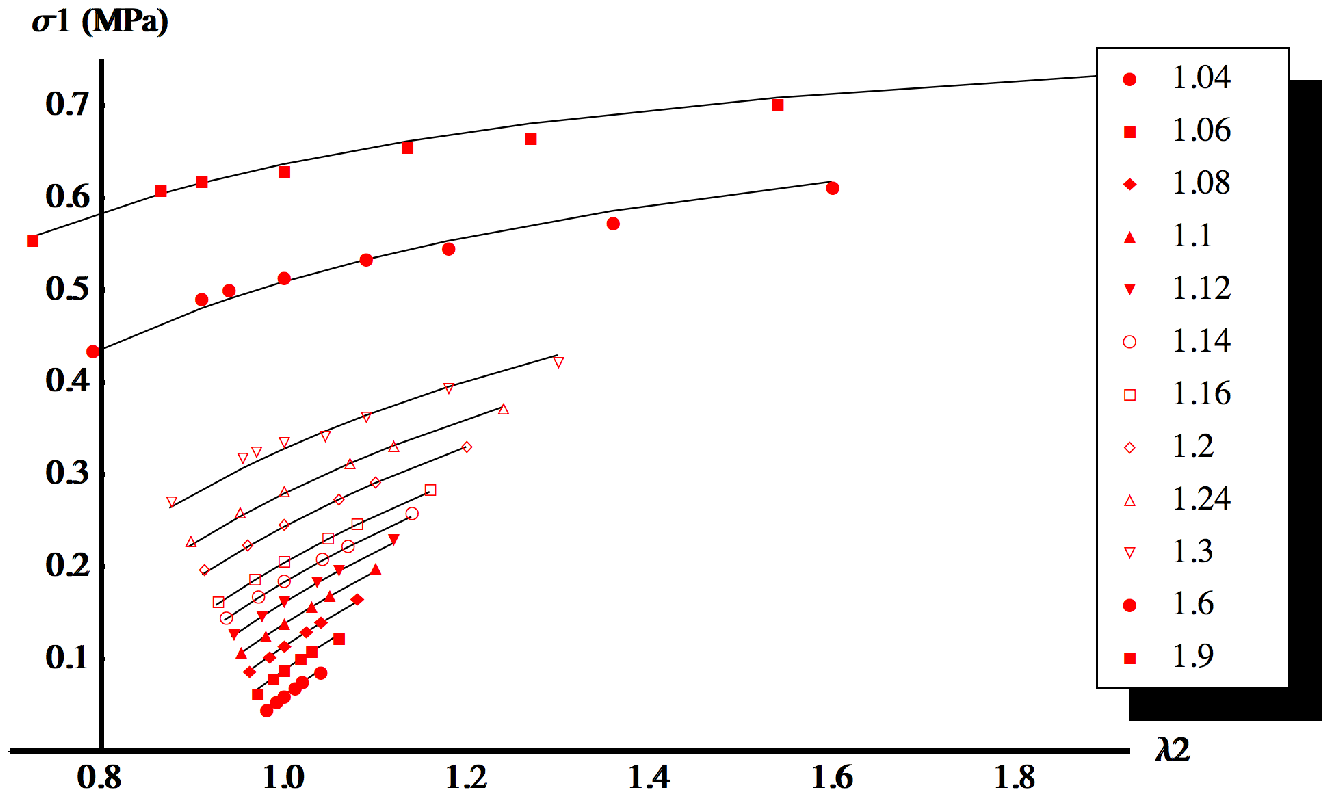}
\caption{
XGR fitting of stress component $\sigma_1 = \partial f/\partial \lambda_1$ for biaxial deformation of Kawabata's sample.  Each symbol corresponds to a distinct value of $\lambda_1$. 
}
%Experimental data and theoretical prediction for $\sigma_1$ as function of $\lambda_2$ for given $\lambda_1$.   The left most points of all branches correspond to uniaxial deformation and were used to extract two parameters $a,b$.    }
\label{biaxial-fitt-Kawabata-1}
\end{center}
%\vspace{-5mm}
\end{figure}

\begin{figure}
\begin{center}
\includegraphics[height=5.5cm]{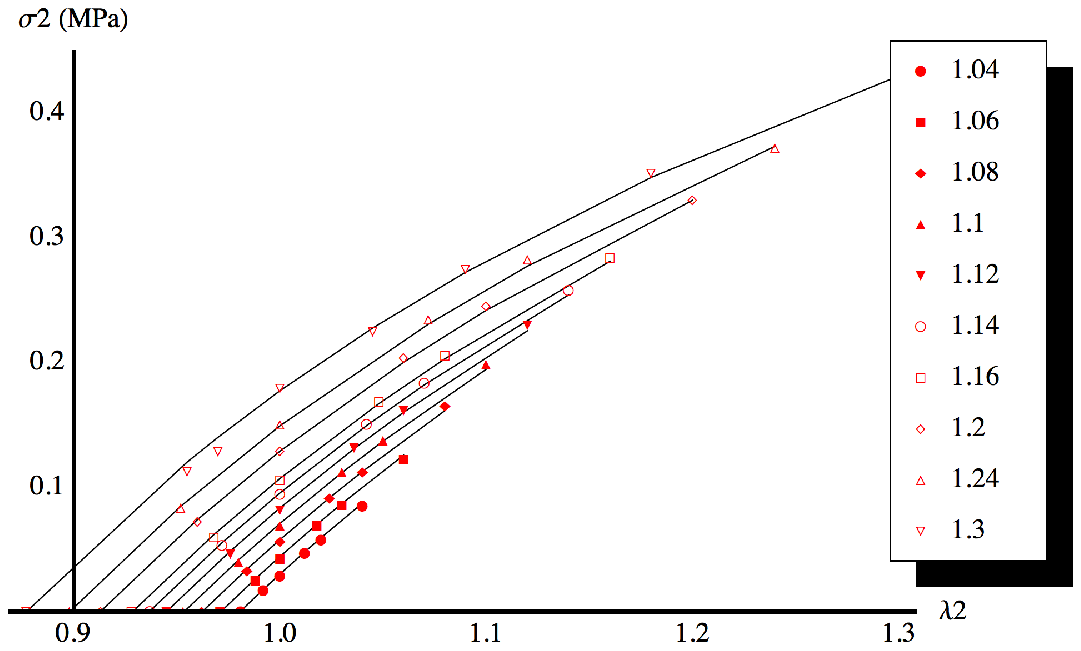}
\caption{
XGR fitting of stress component $\sigma_2 = \partial f/\partial \lambda_2$ for biaxial deformation of Kawabata's sample.  Each symbol corresponds to a distinct value of $\lambda_1$. To improve the visibility of the plot, only data with $\lambda_1 \leq 1.3$ are shown here. 
}
%Experimental data and theoretical prediction for $\sigma_2$ as function of $\lambda_2$ for given $\lambda_1$.   }
\label{biaxial-fitt-Kawabata-2}
\end{center}
%\vspace{-5mm}
\end{figure}

The XGR fit of the uniaxial stress-strain data is shown in Eq.~(\ref{uniaxial-fitt-Kawabata}).  There is a slight upturning trend of the data in the regime $\lambda_1 \geq 2.5$, suggesting that the issue of finite extensibility of polymer chains is coming to play.   This is somewhat expected, since the network being analyzed has a poly-dispersion of chain lengths.  In fact, to avoid dealing with the issue of finite extensibility, we have only used biaxial data with $\lambda_1 \leq 1.9$ for data fitting.  \footnote{As is well understood, the upturning of the stress-strain curve in the large deformation regime is due to the finite extensibility of polymer chains.  There is no fundamental mystery in the understanding of this issue.  XGR does not take into account the finite extensibility of chains, and therefore is not expected to fit the data in the regime of large deformations.  It is therefore not appropriate to compare XGR with the Slip-link theory in this regime.  }The biaxial stress-strain data, as well as their fit using XGR, are shown in Eq.~(\ref{biaxial-fitt-Kawabata-1}) and Eq.~(\ref{biaxial-fitt-Kawabata-2}).   As one can see in Eq.~(\ref{biaxial-fitt-Kawabata-1}), the fitting quality becomes systematically worse as $\lambda_1$ increases.   

\begin{widetext}
\begin{center}
\begin{table}[!hbt]
%\vspace{4mm}
\begin{tabular}{|c|c|c|}
\hline\hline
Theory & Parameters ($MPa$) & Variance ($ 10^9 Pa^2$)
\\ \hline
XGR ($\mu_0 \neq \mu_1$) & $(\mu_0, \mu_1) = (0.294,  0.249)$ & $2.89$ 
\\\hline
XGR ($\mu_0 = \mu_1$) & $\mu_0  = \mu_1 = 0.281$ & $4.07$ 
\\\hline
EV & 
$(C_1,C_2,\eta,\alpha) = (0.22,0.25,0.32, 0.12)$
%$\mu_0 = 0.217238, \mu_1 = 0.251684,  \eta =  0.316602, \alpha =  0.117356$
& $2.13$
\\\hline\end{tabular}
\caption{Best-fitting parameters and variances of Kawabata sample \cite{Kawabata:1981ao}.}
\label{table-Kawabata}
\end{table}
\end{center}
\end{widetext}

\begin{figure}
\begin{center}
\includegraphics[height=5.5cm]{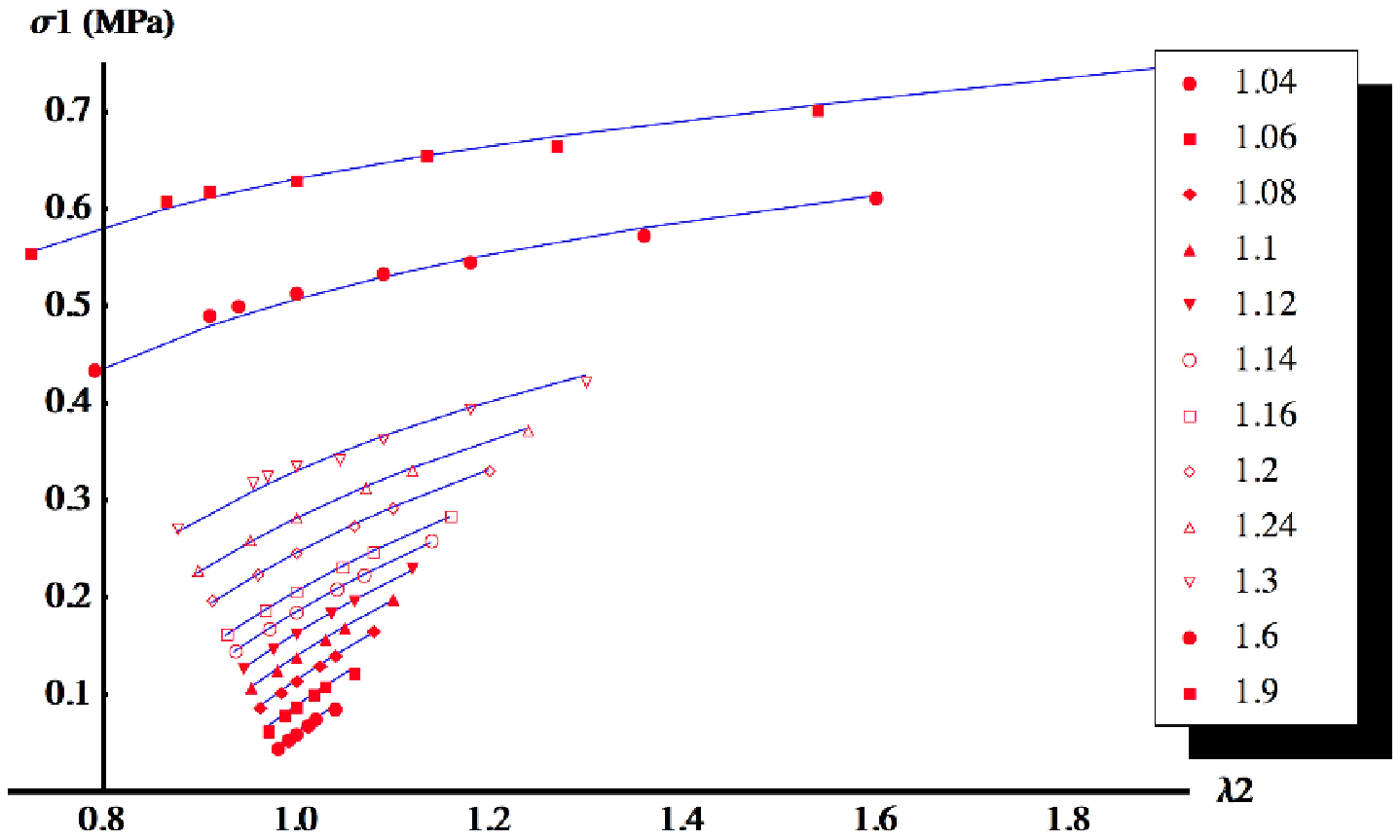}
\caption{ 
EV fitting of stress component $\sigma_1 = \partial f/\partial \lambda_1$ for biaxial deformation of Kawabata's sample.  Each symbol corresponds to a distinct value of $\lambda_1$. 
}
%Fitting of stress component $\sigma_1 = \partial f/\partial \lambda_1$ for biaxial deformation of Kawabata's sample using EV's slip-link theory.  Each symbol corresponds to a distinct value of $\lambda_1$.  }
\label{biaxial-fitt-Kawabata-1-EV}
\end{center}
%\vspace{-5mm}
\end{figure}

\begin{figure}
\begin{center}
\includegraphics[height=5.5cm]{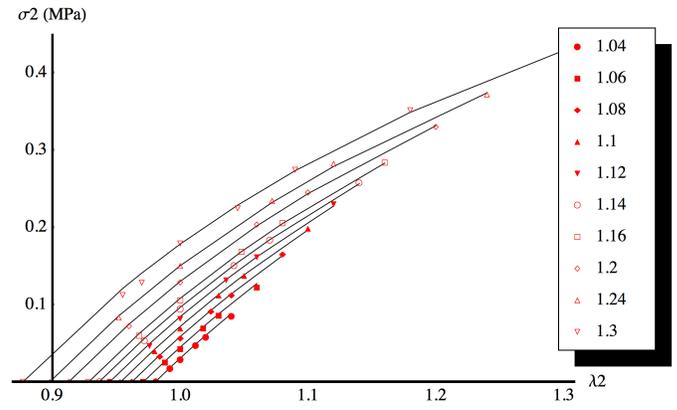}
\caption{
EV fitting of stress component $\sigma_2 = \partial f/\partial \lambda_2$ for biaxial deformation of Kawabata's sample.  Each symbol corresponds to a distinct value of $\lambda_1$. 
To improve the visibility of the plot, only data with $\lambda_1 \leq 1.3$ are shown here. }
%Fitting of stress component $\sigma_2 = \partial f/\partial \lambda_2$ for biaxial deformation of Kawabata's sample using EV's slip-link theory.  Each symbol corresponds to a distinct value of $\lambda_1$.   }
\label{biaxial-fitt-Kawabata-2-EV}
\end{center}
%\vspace{-5mm}
\end{figure}

For comparison, we also show the fitting results of biaxial stress-strain data using EV theory in Eq.~(\ref{biaxial-fitt-Kawabata-1-EV}) and Eq.~(\ref{biaxial-fitt-Kawabata-2-EV}) respectively.  Again no visual difference in fitting quality can be detected between fittings using two theories.  

\section{Discussion and Conclusion}
%{\bf Perspective}   
In this work, we have compared the XGR theory of rubber elasticity~\cite{XGR-rubber} with the slip-link theory using the biaxial stress-strain data of three distinct polymer networks with very different network structures, synthesized by Urayama~\cite{Urayama-biaxial-2}  and Kawabata~\cite{Kawabata:1981ao} respectively.  
With two adjustable parameters, the XGR theory fits three sets of stress-strain data in the whole range of deformation, with comparable quality as the slip-link theory, which has four tunable parameters.  It is probably more remarkable that almost the same fitting quality is achieved by the one parameter version of XGR.  The technical advantages associated with such a one-parameter fitting scheme is rather substantial comparing with various entanglement theories, which typically contain no less than three parameters.  

% From technical point of view, therefore, XGR provides a simple and robust fit to stress-strain data of rubbery materials.  

% We further note that the physics suggested by two theories are substantially different, and even contracting to each other. The XGR theory has not taken into account the effect of finite extensibility of polymer chains.  Therefore the success of XGR in reproducing Urayama's date suggests that the network has not nearly reached its limit of extensibility.   By strong contrast, the parameter for finite extensibility $\alpha$ in slip-link theory, though small, plays essentially important role in the data fitting.  Fitting using this theory with $\alpha$ set to zero is of much worse quality and is not nearly as successful as the XGR theory.   Therefore the fitting of slip-link theory suggests that finite extensibility is an important factor in Urayama's experiment.  Clearly more experiments is needed in order to resolve this issue.   We note, however, the stress-strain curve for uniaxial deformation in Fig.~\ref{uniaxial-fitt-Urayama} shows no sign of up-turning.  This feature therefore contradicts the EV theory. 

The conceptual differences between XGR and entanglement theories are also worth commenting.  Same as the classical theory, almost all entanglement theories of rubber elasticity theory can be categorized as molecular level theories, where a many body problem is reduced to that of a single polymer chain.  By contrast, 
XGR attributes the failure of the classical rubber theory to the entropy associated with the collective long wavelength fluctuations of cross-links, which is manifestly beyond the capacity of all molecular level theories.  Note, however, the strong interplay between long wave length elastic fluctuations and generalized elasticities has been firmly established in the past years, and actually has been one of the main theme of modern condensed matter physics.  Furthermore, XGR demonstrated that the effects of these long wavelength fluctuations should also be detectable using scattering experiments.   By contrast, the effects of entanglement on rubbery elasticity in the regime of large deformation, though are widely believed and have been explored for several decades, have little experimental substantiation beyond fittings of stress-strain data.   
%:
%is still under debate, and so far has not been convincingly established.  
%:
In this perspective, it is rather natural to impose the following question:  what feature in stress-strain data of polymer networks is still unexplained after the effects of long wave-length elastic fluctuations are taken into account?  While the answer to this question suggested by the present work is " probably nothing", careful and extensive comparison using more sets of data is definitely necessary to clarify the situation.  

Perhaps, the most direct experimental test of entanglement theories would be achieved by a study of polymer networks with {\em no} entanglement at all.  Such a network can in principle be prepared by crosslinking a semi-dilute polymer solution at $c^*$ concentration \cite{polymer:deGennes}, where neighboring polymer coils are barely touching each other, but have minimal number of entanglements.  After crosslinking, the solvent should be removed in order to restore the incompressibility.  Any entanglement theory would predict that the elasticity of such a $c^*$ network is described by the classical theory, while according to XGR theory, the elasticity should still be described by the same equation Eq.~(\ref{XGR-free_energy}).  

The current version of XGR theory Eq.~(\ref{XGR-free_energy}) only take into account the first order perturbative correction of elastic fluctuations.  On the other hand, the dimensionless parameter controlling the perturbative series is given by 
$T \, a^{-3}\,\mu$ \cite{XGR-rubber}, which is of order of unity.  Therefore higher order corrections are expected to be quantitatively important, and the good quality of data fitting using first order correction Eq.~(\ref{XGR-free_energy}) may seem fortuitous.  To test the validity of XGR theory beyond leading order, we may numerically simulate the elasticity of an incompressible network in the presence of large elastic fluctuations.  This shall be done in the near future.   

%We hope that this work will stimulate more critical and systematic re-evaluation of   entanglement effects in rubber elasticity.  

%The battle is not finished yet.  We still need  to do more experiments on other aspects of  rubber elasticity.  

\section{Acknowledgement}
The author acknowledges Kenji Urayama for the original data analyzed in this work, as well as for many insightful and interesting comments throughout all stages that lead to this manuscript.  The author also thanks interesting and relevant comments from Ole Hassager, as well as the hospitality of Institute of Physics at Beijing, where part of this work was done. 

\bibliography{/Users/xxing/research/reference-all}

\begin{thebibliography}{10}

\bibitem{XGR-rubber}
Xiangjun Xing, Paul~M. Goldbart, and Leo Radzihovsky.
\newblock Thermal fluctuations and rubber elasticity.
\newblock {\em Physical Review Letters}, 98(7):075502, 2007.

\bibitem{Urayama-biaxial-2}
Kenji Urayama, Takanobu Kawamura, and Shinzo Kohjiya.
\newblock Multiaxial deformations of end-linked poly(dimethylsiloxane)
  networks. 2. experimental tests of molecular entanglement models of rubber
  elasticity.
\newblock {\em Macromolecules}, 34(23):8261--8269, 10 2001.

\bibitem{Kawabata:1981ao}
S.~Kawabata, M.~Matsuda, K.~Tei, and H.~Kawai.
\newblock Experimental survey of the strain energy density function of isoprene
  rubber vulcanizate.
\newblock {\em Macromolecules}, 14(1):154--162, 05 1981.

\bibitem{EL:Treloar}
L.R.G. Treloar.
\newblock {\em The Physics of Rubber Elasticity}.
\newblock Oxford University Press, 1975.

\bibitem{Edwards-tube-1986}
S.~F. Edwards and T.~A. Vilgis.
\newblock The effect of entanglements in rubber elasticity.
\newblock {\em Polymer}, 27(4):483--492, Apr 1986.

\bibitem{Kloczkowski:1995lr}
Andrzej Kloczkowski, James~E. Mark, and Burak Erman.
\newblock A diffused-constraint theory for the elasticity of amorphous polymer
  networks. 1. fundamentals and stress-strain isotherms in elongation.
\newblock {\em Macromolecules}, 28(14):5089--5096, 1995.

\bibitem{Gaylord-Douglas-I}
Richard~J. Gaylord and Jack~F. Douglas.
\newblock Rubber elasticity: a scaling approach.
\newblock {\em Polymer Bulletin}, 18(4):347--354, 10 1987.

\bibitem{Gaylord-Douglas-II}
Richard~J. Gaylord and Jack~F. Douglas.
\newblock The localisation model of rubber elasticity. ii.
\newblock {\em Polymer Bulletin}, 23(5):529--533, 05 1990.

\bibitem{Rubinstein-tube}
M.~Rubinstein and S.~Panyukov.
\newblock Elasticity of polymer networks.
\newblock {\em Macromolecules}, 35:6670, 2002.

\bibitem{james-guth}
Hubert~M. James and Eugene Guth.
\newblock Theory of the elastic properties of rubber.
\newblock {\em The Journal of Chemical Physics}, 11(10):455--481, 1943.

\bibitem{PhysRevE.77.011802}
Rasmus Hansen, Anne~Ladegaard Skov, and Ole Hassager.
\newblock Constitutive equation for polymer networks with phonon fluctuations.
\newblock {\em Phys. Rev. E}, 77(1):011802, Jan 2008.

\bibitem{Urayama-biaxial-1}
Takanobu Kawamura, Kenji Urayama, and Shinzo Kohjiya.
\newblock Multiaxial deformations of end-linked poly(dimethylsiloxane)
  networks. 1. phenomenological approach to strain energy density function.
\newblock {\em Macromolecules}, 34(23):8252--8260, 10 2001.

\bibitem{Urayama-biaxial-4}
Kenji Urayama, Takanobu Kawamura, and Shinzo Kohjiya.
\newblock Multiaxial deformations of end-linked poly(dimethylsiloxane)
  networks. 4. further assessment of the slip-link model for chain-entanglement
  effect on rubber elasticity.
\newblock {\em The Journal of Chemical Physics}, 118(12):5658--5664, 2003.

\bibitem{polymer:deGennes}
P.G. de~Gennes.
\newblock {\em Scaling Concepts in Polymer Physics}.
\newblock Cornell University Press, Itheca, 1979.

\end{thebibliography}

\end{document}